\documentclass[conference]{IEEEtran}
\IEEEoverridecommandlockouts
\usepackage{cite}
\usepackage{amsmath,amssymb,amsfonts}
\usepackage{algorithmic}
\usepackage{graphicx}
\usepackage{textcomp}
\usepackage{xcolor}
\usepackage{multicol}
\usepackage{inputenc}
\usepackage{lipsum}
\usepackage{array,url,kantlipsum}
\usepackage{flushend}
\def\BibTeX{{\rm B\kern-.05em{\sc i\kern-.025em b}\kern-.08em
    T\kern-.1667em\lower.7ex\hbox{E}\kern-.125emX}}
\usepackage{hyperref}

\begin{document}

\twocolumn[{%
  \centering
  \Huge \textbf{Selective Diabetic Retinopathy Screening with Accuracy-Weighted Deep Ensembles and Entropy-Guided Abstention}\\[1em]
  \large \textbf{Jophy Lin}\\[0.25em]
  \href{mailto:jophy.lin@rutgers.edu}{jophy.lin@rutgers.edu}\\[1.25em]
  \normalsize \textit{*Corresponding author}\\[2em]
}]

\begin{abstract}
Diabetic retinopathy (DR), a microvascular complication of diabetes and a leading cause of preventable blindness, is projected to affect more than 130 million individuals worldwide by 2030. Early identification is essential to reduce irreversible vision loss, yet current diagnostic workflows rely on methods such as fundus photography and expert review, which remain costly and resource-intensive. This, combined with DR’s asymptomatic nature, results in its underdiagnosis rate of approximately 25\%. Although convolutional neural networks (CNNs) have demonstrated strong performance in medical imaging tasks, limited interpretability and the absence of uncertainty quantification restrict clinical reliability. Therefore, in this study, a deep ensemble learning framework integrated with uncertainty estimation is introduced to improve robustness, transparency, and scalability in DR detection. The ensemble incorporates seven CNN architectures—ResNet-50, DenseNet-121, MobileNetV3 (Small and Large), and EfficientNet (B0, B2, B3)— whose outputs are fused through an accuracy-weighted majority voting strategy. A probability-weighted entropy metric quantifies prediction uncertainty, enabling low-confidence samples to be excluded or flagged for additional review. Training and validation on 35,000 EyePACS retinal fundus images produced an unfiltered accuracy of 93.70\% (F1 = 0.9376). Uncertainty-filtering later was conducted to remove unconfident samples, resulting in maximum-accuracy of 99.44\% (F1 = 0.9932). The framework shows that uncertainty-aware, accuracy-weighted ensembling improves reliability without hindering performance. With confidence-calibrated outputs and a tunable accuracy–coverage trade-off, it offers a generalizable paradigm for deploying trustworthy AI diagnostics in high-risk care.
\end{abstract}

\section{Introduction}

\subsection{Background and Global Significance}
Diabetic retinopathy (DR) is a progressive eye disease caused by long-term high blood sugar, which damages the small blood vessels in the retina and affects vision. As the leading cause of blindness among working-age adults, it creates major public health and economic challenges. DR develops in two main stages: the nonproliferative stage and the proliferative stage. The nonproliferative stage is signified by microaneurysms, hemorrhages, and leaking vessels, while the proliferative stage is signified by abnormal new blood vessels that grow, which can lead to retinal detachment, a condition where the retina separates from its supporting layer, or permanent loss of vision and blindness \cite{1}.

\begin{figure}[htbp]
  \centering
  \includegraphics[width=0.48\textwidth]{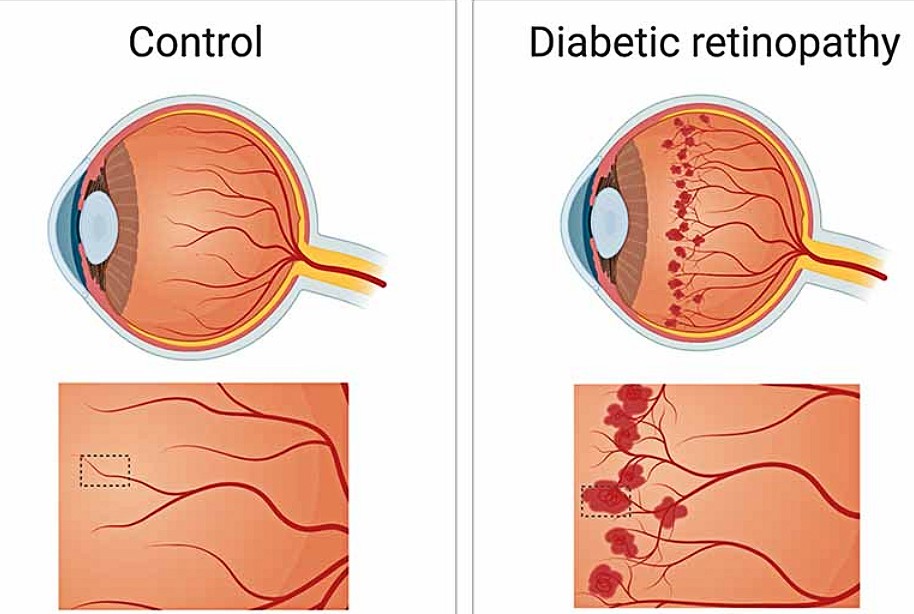}
  \caption{DR progression (pericyte loss → microaneurysms/leakage → neovascularization/scar). 
  Adapted from Fig.~3 in \cite{2}\,(cropped). 
  Licensed under \href{https://creativecommons.org/licenses/by/4.0/}{CC BY 4.0}.}
  \label{fig:dr-illustration}
\end{figure}

By 2030, DR is estimated to affect about 130 million people worldwide \cite{3}. Early detection has been found to be able to prevent up to 70\% of vision loss. However, about one in four (25\%) cases go undiagnosed, and nearly one in five (20\%) individuals have DR at the severe stage, signifying delayed diagnosis, as escalation of the disease can be controlled through early treatment \cite{4}. With treatment costs potentially costing over \$14,000 per patient each year, there’s a clear need for scalable, affordable diagnostic tools that enable early intervention, especially for those in low-resource settings \cite{5}.

\subsection{Limitations of Current Diagnostic Practices}
Typical DR diagnosis currently relies on methods such as dilated fundus examinations, fluorescein angiography, and optical coherence tomography (OCT) \cite{6}. Although clinically effective, these procedures require specialized equipment and expertise that are often unavailable to those in low-resource environments. Additionally, these methods are invasive and can result in various side effects, including nausea and dizziness \cite{7}. Even in advanced healthcare systems, diagnostic variability and human fatigue from long hours or repetitive work can contribute to inconsistent outcomes and delayed treatment. 

Current machine-learning (ML) methods, especially convolutional neural networks (CNNs), have achieved performance high enough to be compared to trained clinicians in large-scale medical image classification tasks \cite{8}. Nevertheless, most generate discrete categorical predictions without expressing confidence levels. This limits interpretability and hinders adoption in clinical workflows and settings where safety and reliability are critical \cite{9}. Furthermore, dataset bias and domain shift can degrade the performance of ML across various populations and imaging devices, showing the importance and need for explicit uncertainty estimation \cite{10}.

\subsection{Need for Uncertainty-Aware Artificial Intelligence}
Uncertainty quantification has become essential for reliable medical-AI deployment. Rather than producing deterministic classifications, models incorporating uncertainty estimation can identify ambiguous inputs for further review, reducing diagnostic risk. Ensemble learning offers a practical solution by combining multiple independent models and interpreting inter-model variability as a measure of predictive confidence \cite{11}. 

The framework presented in this study employs accuracy-weighted ensembling and probability-weighted entropy to quantify uncertainty. The weighting mechanism ensures that stronger base models exert greater influence, improving decision stability while maintaining transparency. This design simultaneously enhances interpretability and predictive robustness, addressing two major obstacles to clinical integration.

\subsection{Objectives and Research Contributions}
The objective of this study is to develop a scalable and interpretable diagnostic architecture for automated DR detection. The primary contributions are as follows:

\begin{itemize}
    \item Construction of a deep ensemble composed of seven CNN architectures (ResNet-50, DenseNet-121, MobileNetV3 Small/Large, and EfficientNet B0/B2/B3) trained on the EyePACS dataset to improve robustness through architectural diversity.
    \item Introduction of a probability-weighted entropy method for sample-level uncertainty estimation and selective prediction.
    \item Demonstration of tunable uncertainty thresholds that permit application-specific trade-offs between accuracy and data coverage.
    \item Empirical validation showing substantial improvement over single-model baselines in both unfiltered and filtered performance metrics.
\end{itemize}

To the best of knowledge, this is the first study to integrate an accuracy-weighted entropy-based uncertainty filter within a multi-architecture ensemble for diabetic retinopathy detection. By coupling interpretability with tunable reliability, the proposed framework advances beyond conventional accuracy-driven systems and establishes a transparent, uncertainty-aware paradigm for clinical AI deployment.

\section{Related Work}

\subsection{Automated Diabetic Retinopathy Detection}
Convolutional neural networks (CNNs) have achieved notable success in automated diabetic retinopathy (DR) classification tasks, particularly on large-scale datasets such as EyePACS and Messidor. Early competitive models, including InceptionV3-based frameworks and EfficientNet architectures, demonstrated that deep feature extraction can match or surpass clinician-level accuracy in fundus image analysis~\cite{12, 13}. Commercial systems such as IDx-DR have further validated the clinical feasibility of AI-assisted DR screening by achieving FDA approval for autonomous operation~\cite{14}. However, despite high accuracy, these systems often lack transparency and do not quantify uncertainty, limiting their adoption in high-risk medical settings.

\subsection{Ensemble Deep Learning in Medical Imaging}
Ensemble learning has emerged as a robust approach for improving diagnostic accuracy and mitigating overfitting in medical image analysis. For example, Heisler \textit{et al.} demonstrated that ensemble CNN architectures achieved higher accuracy than individual networks in classifying referable diabetic retinopathy from OCT-A datasets~\cite{15}. Likewise, Ai \textit{et al.} introduced the “DR-IIXRN” deep ensemble framework—comprising Inception V3, Inception-ResNet V2, Xception, ResNeXt101, and NASNetLarge—with weighted voting, specifically applied to DR detection from fundus images~\cite{16}. While these studies achieved strong performance, most prior ensemble systems remain deterministic and lack explicit uncertainty quantification. In contrast, the framework advances existing work by incorporating a probabilistic confidence metric that enables selective filtering of low-confidence predictions.

\subsection{Uncertainty Estimation in Medical AI}
Quantifying predictive uncertainty is increasingly recognized as essential for safe AI deployment in healthcare. Bayesian convolutional neural networks (CNNs)~\cite{17} and Monte Carlo dropout methods~\cite{18} have been proposed to estimate epistemic and aleatoric uncertainty, while evidential deep learning frameworks~\cite{19} offer single-pass alternatives. Despite their theoretical promise, these techniques are computationally intensive and often challenging to interpret in clinical workflows. Moreover, few have been evaluated specifically in ophthalmic imaging or integrated with ensemble-based architectures.

\subsection{Gap and Contribution}
While prior research has addressed DR classification and explored ensemble methods, limited attention has been given to integrating uncertainty quantification within ensemble frameworks for medical imaging. The proposed accuracy-weighted entropy ensemble bridges this gap by combining multiple CNN architectures with a probabilistic confidence metric, enabling selective predictions based on tunable uncertainty thresholds. This approach enhances both diagnostic reliability and interpretability, providing a scalable and transparent foundation for trustworthy AI in diabetic retinopathy detection.

\section{Methodology}

\subsection{Overview}
Ensemble learning is a ML paradigm that combines predictions from multiple models to produce a single, more accurate result. This approach leverages the complementary strengths of individual learners, thereby generating more robust and reliable outcomes. By incorporating strategies such as majority voting and techniques including subsampling and data augmentation, ensemble learning can mitigate weaknesses inherent to individual models while enhancing generalization capability \cite{20}. 

\begin{figure}[htbp]
\centerline{\includegraphics[width=0.48\textwidth]{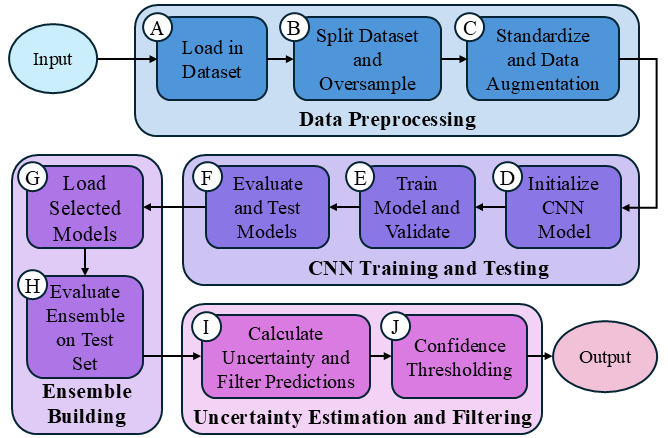}}
\caption{Sequential Workflow of the Proposed Methodology.}
\label{fig:workflow}
\end{figure}

In this framework, ensemble learning not only performs classification but also integrates a novel uncertainty-estimation mechanism for evaluating prediction confidence. This integration enhances both the interpretability and reliability of the diagnostic system. The sequential workflow of the proposed methodology is summarized in Fig.~\ref{fig:workflow}, with detailed subsections provided below.

\subsection{Data Preprocessing}
\begin{figure}[htbp]
\centerline{\includegraphics[width=0.48\textwidth]{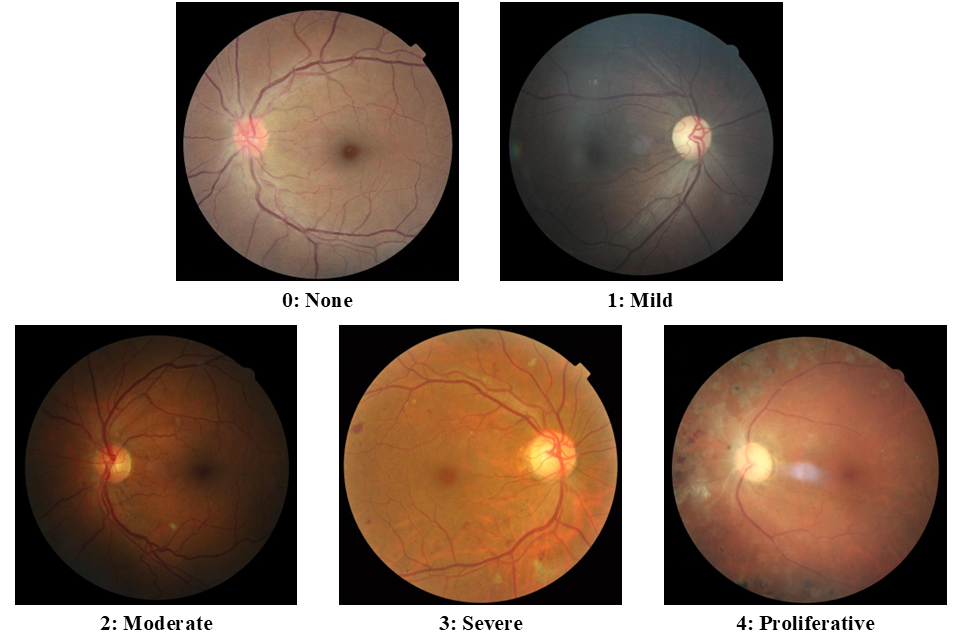}}
\caption{Representative Data Samples at Various DR Severity Levels}
\label{fig:samples}
\end{figure}

The EyePACS dataset, a publicly available collection of over 35,000 retinal fundus images labeled for diabetic-retinopathy (DR) severity on a five-level scale (0–4), was utilized \cite{21}. Labels were defined as follows: 0 indicates no DR, whereas labels 1 through 4 represent progressively severe stages ranging from mild to proliferative DR. Representative examples for each category are shown in Fig.~\ref{fig:samples}, illustrating characteristic retinal changes associated with disease progression. Images labeled as 0 display healthy retinal morphology, while those labeled as 4 exhibit extensive proliferation.

\subsection{Data Conversion and Splitting}
To align the classification task with current clinical priorities, the original five-class DR severity scale was converted into a binary format. Images labeled as~0 were categorized as ``No~DR,'' whereas labels~1–4 were grouped as ``Presence~of~DR.'' This design was decided due to the most critical challenge in DR screening: accurately identifying whether a patient has DR in the first place, given its asymptomatic nature. The binary framework mirrors real-world clinical screening, where the primary concern is distinguishing affected individuals from healthy patients to enable timely intervention. The dataset was divided into training, validation, and testing subsets containing approximately 26,000, 5,000, and 5,000 images, respectively. Stratified sampling preserved class balance across all subsets, ensuring consistent evaluation and minimizing sampling bias.

\subsection{Addressing Class Imbalance}
With the dataset, a class imbalance was observed, with non-DR images substantially outnumbering DR-positive samples. To counteract this, oversampling was applied to the minority class in the training subset, thereby achieving approximate class balance. This enhanced sensitivity to DR-positive instances and reduced prediction bias.

\subsection{Data Augmentation}
To improve generalization and reduce overfitting, data augmentation was employed using geometric and photometric transformations. Random rotations, horizontal and vertical flips, and brightness adjustments were applied to expand the diversity of the training data and help the models generalize better. Images were cropped to a square aspect ratio and resized to $224\times224$~pixels for compatibility with CNN architectures. Each subset was independently standardized to zero mean and unit variance, a procedure that accelerates convergence by maintaining consistent input scales during optimization.

\subsection{CNN Model Training and Testing}
Given the complexity of DR presentation and the subtle gradations between stages, a diverse set of CNNs was selected to capture heterogeneous retinal features. The ensemble consisted of ResNet-50, DenseNet-121, MobileNetV3 (Small and Large), and EfficientNet variants (B0, B2, B3). Each model was fine-tuned for binary classification using transfer learning from ImageNet-pretrained weights.

\subsubsection{Model Selection}
Each CNN architecture contributed distinct advantages to the ensemble. 

ResNet-50, a 50-layer residual network, employs skip connections to alleviate vanishing-gradient issues and excels in learning hierarchical features, thereby facilitating the detection of microaneurysms and subtle hemorrhages in early DR \cite{22}. 

DenseNet-121 enhances feature reuse by connecting every layer to all preceding layers, enabling simultaneous capture of local and global features and minimizing redundancy \cite{23}. 

MobileNetV3 architectures are optimized for computational efficiency through depthwise separable convolutions and squeeze-and-excitation modules, providing robustness in real-time or resource-constrained environments \cite{24}. 

EfficientNet models apply compound scaling to balance depth, width, and resolution, enabling high accuracy with optimal efficiency. The larger variants, such as EfficientNetB3, effectively capture high-resolution patterns including neovascularization and large hemorrhages, while smaller variants (B0, B2) deliver lightweight adaptability \cite{25}.

\subsubsection{Leveraging Model Diversity}
The diversity among architectures ensures complementary coverage of retinal pathology. For instance, while ResNet-50 effectively models hierarchical relationships, it may overlook finer-grained lesions that DenseNet-121 captures. MobileNetV3 adds computational efficiency, and EfficientNet’s compound scaling facilitates high-resolution analysis. Integrating models with varied feature hierarchies enables the ensemble to perform robustly across the full spectrum of DR severity, from subtle early indicators to advanced pathological states.

\begin{figure}[htbp]
\centerline{\includegraphics[width=0.48\textwidth]{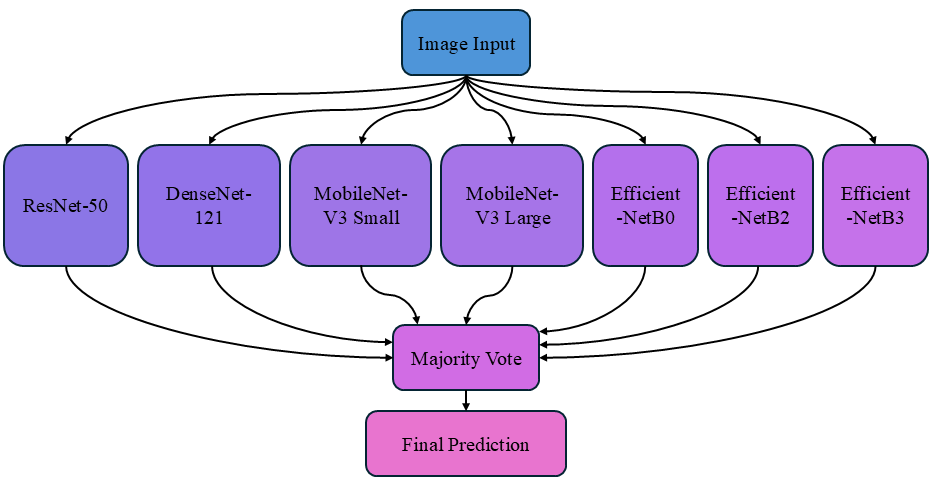}}
\caption{Architecture of the Ensemble Learning Framework}
\label{fig:ensemble}
\end{figure}

\subsubsection{Training Configuration and Checkpoint Selection}
All models were initialized with pretrained weights and fine-tuned using the Binary Cross-Entropy with Logits loss function, which combines sigmoid activation and loss computation into a single numerically stable operation. The Adam optimizer was employed with a learning rate of 0.001 to balance convergence speed and stability. A batch size of 16 was used to optimize computational efficiency. 

Training durations were 100~epochs for ResNet-50 and DenseNet-121, and 50~epochs for MobileNetV3 and EfficientNet models, with early stopping applied to prevent overfitting. Performance metrics, including loss, accuracy, and F1 score, were recorded on both training and validation sets. Model checkpoints were saved every 10 epochs, and the best-performing models (highest validation F1 and accuracy) were selected for ensemble aggregation.

\subsection{Ensemble Building}
The ensemble was constructed by aggregating outputs from all trained CNNs. Each model’s raw logits were converted to probabilities using the sigmoid activation function:
\begin{equation}
P_i = \sigma(z_i) = \frac{1}{1 + e^{-z_i}},
\end{equation}
where $z_i$ represents the output logit of model~$i$ and $\sigma$ denotes the sigmoid function. 

Two aggregation strategies were evaluated: standard majority voting and accuracy-weighted majority voting. In standard majority voting, binary predictions from each model were tallied, and the class with the highest frequency was assigned as the ensemble prediction. In contrast, accuracy-weighted voting assigned greater influence to models with superior validation performance. 

The normalized accuracy weight for model~$i$ was defined as:
\begin{equation}
W_i = \frac{a_i}{\sum_{k=1}^{n} a_k},
\end{equation}
where $a_i$ represents the validation accuracy of model~$i$ and $n$ denotes the total number of models. 

The final ensemble prediction $\hat{y}$ was computed as the weighted sum of individual model outputs:
\begin{equation}
\hat{y} = \sum_{i=1}^{n} W_i \, v_i,
\end{equation}
where $v_i$ denotes the predicted probability from model~$i$. Substituting the definition of $W_i$ into this equation yields:
\begin{equation}
\hat{y} = 
\frac{\sum_{i=1}^{n} a_i \, v_i}{\sum_{i=1}^{n} a_i},
\end{equation}
which matches the formulation illustrated in (4). This expression demonstrates that the ensemble prediction is a normalized, accuracy-weighted average of individual model probabilities, ensuring that higher-performing models contribute more strongly to the final output. Weighted voting achieved the highest unfiltered accuracy and was selected for the final implementation.

The final ensemble was evaluated on the held-out test set using accuracy, F1 score, precision, and recall prior to uncertainty filtering.

\subsection{Uncertainty Estimation and Thresholding}
To quantify prediction confidence and enhance diagnostic reliability, uncertainty estimation was applied to ensemble outputs. The dispersion of probabilities across models reflects prediction confidence: lower variance corresponds to higher certainty. A tunable entropy-based threshold was introduced to discard predictions below a confidence cutoff, thus ensuring that only high-certainty results contribute to the final diagnostic output.

Samples failing to meet the confidence threshold are excluded or flagged for secondary clinical review. This selective prediction mechanism aligns with medical requirements for reliability in high-stakes decision-making.

\subsubsection{Uncertainty Estimation Metrics}
Multiple uncertainty metrics were evaluated for effectiveness. The first, unweighted standard deviation, computes the standard deviation of model probabilities without weighting, providing a basic measure of variance. The second, unweighted binary-output entropy, measures ensemble disagreement by treating all models equally. The third, accuracy-weighted binary-output entropy, incorporates model accuracy into entropy computation, granting higher influence to reliable models. The fourth and final metric, accuracy-weighted probability entropy, calculates entropy directly from weighted probabilities, offering a refined measure of uncertainty that accounts for both probability distribution and model reliability. 

Entropy, or Shannon information entropy, is defined as \cite{26}:
\begin{equation}
H(\hat{p}) = -\,\hat{p}\,\log \hat{p} - (1-\hat{p})\,\log(1-\hat{p}).
\label{eq:shannon-binary}
\end{equation}

where $P_i$ denotes the probability of the positive class. Entropy values close to~0 indicate high agreement among models, whereas higher entropy signifies uncertainty and inter-model disagreement.

\subsubsection{Performance Assessment}
Each uncertainty-estimation approach was evaluated using unfiltered and filtered metrics, including accuracy and F1 score. ``Filtered'' metrics denote evaluations after exclusion of uncertain samples. 

\section{Results}

\subsection{Individual Model Performance}
The performance of each individual CNN model on the held-out test set is summarized in Table~\ref{tab:individual}. Evaluation metrics include accuracy, F1~score, precision, and recall. The variability across architectures demonstrates the complementary nature of the selected models and supports their integration into an ensemble framework. Among all models, EfficientNetB3 achieved the highest overall performance across all metrics.

\begin{table}[htbp]
\centering
\caption{Performance of Individual Models on the Test Set}
\label{tab:individual}
\begin{tabular}{lcccc}
\hline
\textbf{Model} & \textbf{Accuracy} & \textbf{F1} & \textbf{Precision} & \textbf{Recall} \\
\hline
ResNet-50 & 0.8081 & 0.8081 & 0.8082 & 0.8081 \\
DenseNet-121 & 0.8157 & 0.8212 & 0.8267 & 0.8157 \\
MobileNetV3-Small & 0.8223 & 0.8300 & 0.8378 & 0.8223 \\
MobileNetV3-Large & 0.8516 & 0.8536 & 0.8557 & 0.8516 \\
EfficientNetB0 & 0.8901 & 0.8904 & 0.8908 & 0.8901 \\
EfficientNetB2 & 0.9034 & 0.9034 & 0.9035 & 0.9034 \\
EfficientNetB3 & \textbf{0.9088} & \textbf{0.9089} & \textbf{0.9090} & \textbf{0.9088} \\
\hline
\end{tabular}
\end{table}

\begin{figure}[htbp]
\centerline{\includegraphics[width=0.39\textwidth]{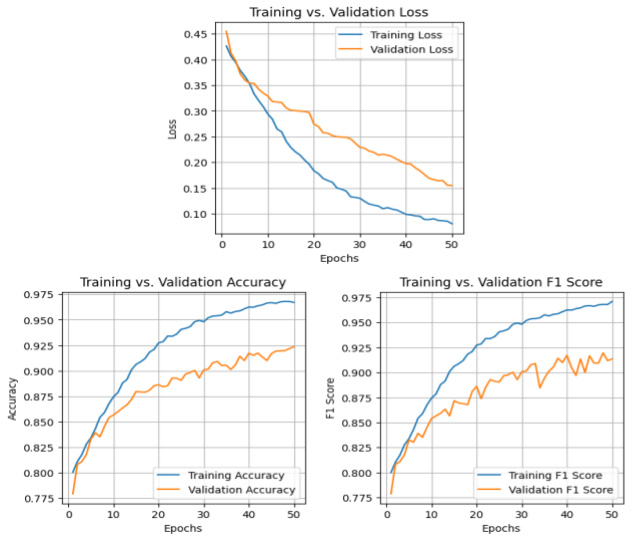}}
\caption{Training curves for EfficientNetB3 (Loss, Accuracy, F1 Score)}
\label{fig:learningcurve}
\end{figure}

EfficientNetB3 exhibited consistent loss reduction during training without indications of overfitting, as illustrated in Fig.~\ref{fig:learningcurve}. Both accuracy and F1~score increased steadily across epochs, underscoring its robustness for the classification task. Similar learning trends were observed in the remaining architectures, confirming their suitability for ensemble integration.

\subsection{Ensemble Performance}
The performance comparison between standard and accuracy-weighted majority-voting strategies is summarized in Table~\ref{tab:voting}. Empirical evaluation indicated that accuracy-weighted voting consistently produced superior outcomes across all evaluation metrics, including accuracy, precision, recall, and F1~score. This improvement demonstrates that assigning greater influence to higher-performing models yields more reliable consensus predictions and mitigates the effects of weaker individual learners. Using this strategy, the final ensemble achieved an unfiltered accuracy of 93.70\% and an F1~score of~0.9376, outperforming the standard majority-voting approach. Owing to its enhanced stability and predictive consistency, the accuracy-weighted configuration was selected for all subsequent analyses and uncertainty-estimation experiments.

\begin{table}[htbp]
\centering
\caption{Comparison of Voting Strategies for Ensemble Aggregation}
\label{tab:voting}
\begin{tabular}{lcc}
\hline
\textbf{Voting Strategy} & \textbf{Accuracy} & \textbf{F1~Score} \\
\hline
Standard Majority & 0.9359 & 0.9359 \\
Accuracy-Weighted & \textbf{0.9370} & \textbf{0.9376} \\
\hline
\end{tabular}
\end{table}

\subsection{Uncertainty Estimation and Thresholding}
\begin{figure}[htbp]
\centerline{\includegraphics[width=0.48\textwidth]{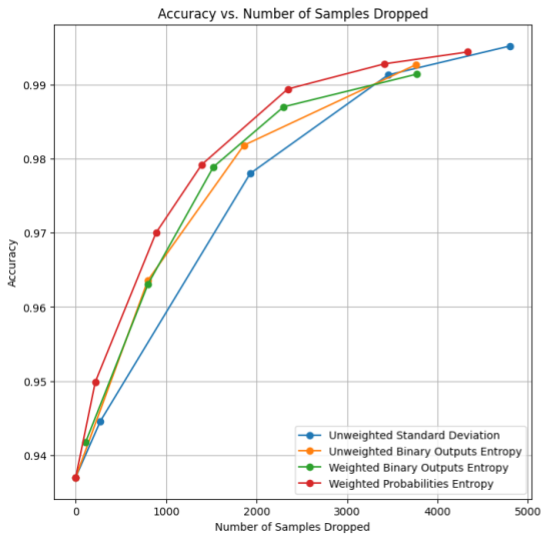}}
\caption{Comparison of Uncertainty-Estimation Metrics and Their Effect on Accuracy versus Percentage of Samples Discarded}
\label{fig:uncertainty}
\end{figure}

Multiple uncertainty-estimation metrics were evaluated to assess their impact on ensemble reliability and diagnostic confidence. Among the tested approaches, accuracy-weighted entropy derived from continuous probability distributions demonstrated the most stable and discriminative performance, achieving the highest area under the curve (AUC), as illustrated in Fig.~\ref{fig:uncertainty}. This metric effectively captured inter-model disagreement by accounting for both probability dispersion and model reliability, providing a robust quantitative indicator of predictive confidence. As a result, accuracy-weighted entropy established a consistent and interpretable basis for identifying low-confidence predictions, which could then be flagged for further review or excluded from automated decision-making to enhance the overall dependability of the system.

Applying an entropy threshold of 0.38 yielded a filtered ensemble accuracy of 99.44\% and an F1 score of 0.9932. Table III summarizes the trade-offs between uncertainty thresholds, data retention, and filtered performance. Lower thresholds remove more ambiguous samples, resulting in higher accuracy but reduced dataset coverage

\begin{table}[htbp]
\centering
\caption{Effect of Uncertainty Thresholds on Data Retention and Accuracy}
\label{tab:threshold}
\begin{tabular}{ccccc}
\hline
\textbf{Threshold} & \textbf{Samples} & \textbf{\%~Dropped} & \textbf{Acc.} & \textbf{F1} \\
\hline
0.38 & 3{,}633 & 69.2 & \textbf{0.9944} & \textbf{0.9932} \\
0.50 & 2{,}342 & 44.6 & 0.9900 & 0.9910 \\
0.60 & 1{,}402 & 26.7 & 0.9789 & 0.9802 \\
0.65 & 891 & 17.0 & 0.9698 & 0.9710 \\
0.69 & 222 & 4.2 & 0.9489 & 0.9498 \\
0.70 & 0 & 0.0 & 0.9370 & 0.9376 \\
\hline
\end{tabular}
\end{table}

\section{Discussion}

\subsection{Interpretation of Results}
The ensemble learning framework augmented with uncertainty estimation enhanced both diagnostic reliability and predictive accuracy for diabetic-retinopathy (DR) detection. Results confirm that the ensemble significantly outperformed individual CNN models across all evaluated metrics. The strongest single model, EfficientNetB3, achieved 90.88\% accuracy and an F1~score of~0.9089, whereas the accuracy-weighted ensemble produced an accuracy of~93.70\% and an F1~score of~0.9376. These results demonstrate the benefit of combining multiple architectures to exploit complementary strengths and reduce the limitations inherent in single-model systems.

The application of uncertainty estimation further improved reliability by enabling selective exclusion of low-confidence predictions. At an entropy threshold of~0.38, the filtered ensemble reached near-perfect accuracy (99.44\%) and an F1~score of~99.32\%. However, this improvement required excluding 69.2\% of test samples, illustrating a trade-off between diagnostic certainty and data coverage. In medical contexts, such balance is critical: retaining too many uncertain cases increases risk of misclassification, whereas discarding excessive samples may delay timely intervention.

\subsection{Key Findings}
Individual model performances revealed substantial variability, underscoring the necessity of ensembling. EfficientNetB3 delivered the strongest baseline due to its compound scaling, which optimizes network depth, width, and resolution, thereby enabling precise identification of fine-grained retinal abnormalities. Nonetheless, reliance on a single model proved insufficient for addressing the full spectrum of DR variability. Integration of architectures such as DenseNet-121 and ResNet-50 provided complementary feature extraction, enhancing the ensemble’s robustness.

Comparison of aggregation strategies confirmed that accuracy-weighted voting slightly outperformed standard majority voting. Although the observed improvement (0.9359 → 0.9370) was modest, such refinements are clinically meaningful in high-stakes diagnostic applications. Even marginal gains can substantially affect patient outcomes when scaled to population-level screening.

Accuracy-weighted probability entropy emerged as the most effective uncertainty metric, achieving the highest AUC and delivering the strongest trade-off between accuracy and coverage. As indicated in Table~\ref{tab:threshold}, performance declines gradually as the entropy threshold increases, providing flexibility for clinical tuning. Lower thresholds yield extremely high reliability at the cost of coverage, whereas higher thresholds retain more samples with slightly reduced accuracy.

\subsection{Significance of Findings}
The study demonstrates that ensemble learning can address diagnostic variability caused by differences in imaging conditions, disease progression, and patient demographics. The integration of diverse architectures enables comprehensive feature representation: residual connections in ResNet-50 facilitate hierarchical learning, dense connectivity in DenseNet-121 promotes feature reuse, and EfficientNet’s compound scaling captures fine detail. Together, these characteristics yield a more stable and interpretable diagnostic framework.

Incorporating uncertainty estimation introduces a critical safeguard for clinical adoption. By flagging ambiguous cases, the system supports human oversight and reduces potential diagnostic risk. Tunable confidence thresholds enable adaptation to diverse workflows: low thresholds may benefit early-screening programs prioritizing sensitivity, while high thresholds suit confirmatory diagnostics emphasizing specificity. Such flexibility enhances applicability across healthcare contexts, particularly in regions with limited ophthalmologic resources.

\subsection{Limitations and Future Directions}
Several limitations remain. Reliance on the EyePACS dataset may restrict generalizability, as this dataset does not fully represent global imaging variability. External validation across independent datasets is necessary to confirm robustness. The binary classification approach simplifies DR detection but omits granularity among severity grades; future work should extend the framework to multi-class classification for improved clinical utility.

Although uncertainty filtering increased diagnostic precision, it also excluded a substantial portion of data, potentially limiting practical deployment where comprehensive screening is desired. Incorporating semi-supervised or Bayesian approaches may yield more balanced uncertainty estimation. Additionally, future implementations should focus on real-time integration with ophthalmic workflows and visualization tools that communicate uncertainty effectively to clinicians. Longitudinal evaluations in clinical environments are recommended to assess sustained reliability and scalability.

\section{Conclusion}
A novel ensemble learning framework incorporating uncertainty estimation was presented for diabetic-retinopathy (DR) detection. Beyond conventional emphasis on predictive accuracy, the proposed framework underscores the transformative role of interpretable artificial intelligence in clinical diagnostics. By integrating diverse convolutional neural-network architectures within an accuracy-weighted entropy mechanism, the framework demonstrates that ensemble-based approaches can exceed traditional diagnostic paradigms in both reliability and adaptability.

The system’s capacity to generate actionable uncertainty metrics marks a shift from opaque “black-box’’ outputs toward transparent, confidence-aware decision making. Such interpretability strengthens clinical trust and supports safe deployment in real-world healthcare environments, including resource-limited regions. The framework thus provides not only enhanced diagnostic performance but also a scalable foundation for trustworthy AI integration in medicine.

Future work should extend the binary-classification design to a multi-class formulation to distinguish among disease severities, incorporate cross-dataset generalization, and integrate multimodal clinical data. These refinements would position the framework as a versatile platform applicable to other medical-imaging challenges, such as early detection of additional pathologies or longitudinal patient monitoring.

The findings emphasize the dual value of artificial intelligence in healthcare: advancing data-driven precision while reinforcing clinical usability. The demonstrated balance between analytical accuracy and interpretability establishes a pathway for next-generation diagnostic systems aimed at improving accessibility, reducing misdiagnosis, and enhancing trust in high-stakes medical decision making.

\section{Reproducibility and Implementation Details}
All experiments utilized the EyePACS dataset (2015 Kaggle competition). Training was performed in Google Colab’s high-RAM environment with the Tesla V100 GPU. Models were optimized using Adam ($\text{lr}=0.001$) for up to 100~epochs with early stopping based on validation loss. Code and preprocessing scripts are publicly available at https://github.com/jophy2467.

\section{Limitations and Ethical Considerations}
This research was conducted using publicly available retinal images and does not involve patient-identifiable data. The framework has not been validated in clinical trials, and generalization to diverse imaging devices or populations is not guaranteed. Future research will include cross-dataset validation, fairness assessments across demographics, and integration into ophthalmic workflows to ensure safe clinical deployment.

\end{document}